
\newbox\leftpage \newdimen\fullhsize \newdimen\hstitle \newdimen\hsbody
\tolerance=1000\hfuzz=2pt
\magnification=1200\baselineskip=16pt plus 2pt minus 1pt
\parindent=.31truein
\hoffset=0truein
\voffset=.2truein
\hsbody=\hsize \hstitle=\hsize 
%
\catcode`\@=11 
\newcount\yearltd\yearltd=\year\advance\yearltd by -1900

\def\Title#1#2{\nopagenumbers\abstractfont\hsize=\hstitle\rightline{#1}%
\vskip 1in\centerline{\titlefont #2}\abstractfont\vskip .5in\pageno=0}
\def\Date#1{\vfill\leftline{#1}\tenpoint\supereject\global\hsize=\hsbody%
\footline={\hss\tenrm\folio\hss}}
\def\draftmode{\def\draftdate{{\rm preliminary draft:
\number\month/\number\day/\number\yearltd\ \ \hourmin}}%
\headline={\hfil\draftdate}\writelabels\baselineskip=20pt plus 2pt minus 2pt
{\count255=\time\divide\count255 by 60 \xdef\hourmin{\number\count255}
        \multiply\count255 by-60\advance\count255 by\time
   \xdef\hourmin{\hourmin:\ifnum\count255<10 0\fi\the\count255}}}

\def\nolabels{\def\eqnlabel##1{}\def\eqlabel##1{}\def\reflabel##1{}}
\def\writelabels{\def\eqnlabel##1{%
{\escapechar=` \hfill\rlap{\hskip.09in\string##1}}}%
\def\eqlabel##1{{\escapechar=` \rlap{\hskip.09in\string##1}}}%
\def\reflabel##1{\noexpand\llap{\string\string\string##1\hskip.31in}}}
\nolabels
%
\global\newcount\secno \global\secno=0
\global\newcount\meqno \global\meqno=1
\def\newsec#1{\global\advance\secno by1
\xdef\secsym{\the\secno.}\global\meqno=1
\bigbreak\bigskip
\noindent{\bf\the\secno. #1}\par\nobreak\medskip\nobreak}
\xdef\secsym{}
\def\appendix#1#2{\global\meqno=1\xdef\secsym{\hbox{#1.}}\bigbreak\bigskip
\noindent{\bf Appendix #1. #2}\par\nobreak\medskip\nobreak}
%
%
\def\eqnn#1{\xdef #1{(\secsym\the\meqno)}%
\global\advance\meqno by1\eqnlabel#1}
\def\eqna#1{\xdef #1##1{\hbox{$(\secsym\the\meqno##1)$}}%
\global\advance\meqno by1\eqnlabel{#1$\{\}$}}
\def\eqn#1#2{\xdef #1{(\secsym\the\meqno)}\global\advance\meqno by1%
$$#2\eqno#1\eqlabel#1$$}
%
\newskip\footskip\footskip14pt plus 1pt minus 1pt 
\def\f@@t{\baselineskip\footskip\bgroup\aftergroup\@foot\let\next}
\setbox\strutbox=\hbox{\vrule height9.5pt depth4.5pt width0pt}
\global\newcount\ftno \global\ftno=0
\def\foot{\global\advance\ftno by1\footnote{$^{\the\ftno}$}}
%
%
\global\newcount\refno \global\refno=1
\newwrite\rfile
\def\ref{[\the\refno]\nref}
\def\nref#1{\xdef#1{[\the\refno]}\ifnum\refno=1\immediate
\openout\rfile=refs.tmp\fi\global\advance\refno by1\chardef\wfile=\rfile
\immediate\write\rfile{\noexpand\item{#1\ }\reflabel{#1}\pctsign}\findarg}
\def\findarg#1#{\begingroup\obeylines\newlinechar=`\^^M\pass@rg}
{\obeylines\gdef\pass@rg#1{\writ@line\relax #1^^M\hbox{}^^M}%
\gdef\writ@line#1^^M{\expandafter\toks0\expandafter{\striprel@x #1}%
\edef\next{\the\toks0}\ifx\next\em@rk\let\next=\endgroup\else\ifx\next\empty%
\else\immediate\write\wfile{\the\toks0}\fi\let\next=\writ@line\fi\next\relax}}
\def\striprel@x#1{} \def\em@rk{\hbox{}} {\catcode`\%=12\xdef\pctsign{
\def\semi{;\hfil\break}
\def\addref#1{\immediate\write\rfile{\noexpand\item{}#1}} 
\def\listrefs{\vfill\eject\immediate\closeout\rfile
\baselineskip=14pt\centerline{{\bf References}}\bigskip{\frenchspacing%
\escapechar=` \input refs.tmp\vfill\eject}\nonfrenchspacing}
\def\startrefs#1{\immediate\openout\rfile=refs.tmp\refno=#1}
\def\figures{\centerline{{\bf Figure Captions}}\medskip\parindent=40pt}
\def\fig#1#2{\medskip\item{Fig.~#1:  }#2}
\catcode`\@=12 
%
\font\titlerm=cmr10 scaled\magstep3 \font\titlerms=cmr7 scaled\magstep3
\font\titlermss=cmr5 scaled\magstep3 \font\titlei=cmmi10 scaled\magstep3
\font\titleis=cmmi7 scaled\magstep3 \font\titleiss=cmmi5 scaled\magstep3
\font\titlesy=cmsy10 scaled\magstep3  \font\titlesys=cmsy7 scaled\magstep3
\font\titlesyss=cmsy5 scaled\magstep3 \font\titleit=cmti10 scaled\magstep3
\skewchar\titlei='177 \skewchar\titleis='177 \skewchar\titleiss='177
\skewchar\titlesy='60 \skewchar\titlesys='60 \skewchar\titlesyss='60
\def\titlefont{\def\rm{\fam0\titlerm}
\textfont0=\titlerm \scriptfont0=\titlerms \scriptscriptfont0=\titlermss
\textfont1=\titlei \scriptfont1=\titleis \scriptscriptfont1=\titleiss
\textfont2=\titlesy \scriptfont2=\titlesys \scriptscriptfont2=\titlesyss
\textfont\itfam=\titleit \def\it{\fam\itfam\titleit} \rm}
\def\abstractfont{\tenpoint}

\def\tenpoint{\def\rm{\fam0\tenrm}
\textfont0=\tenrm \scriptfont0=\sevenrm \scriptscriptfont0=\fiverm
\textfont1=\teni  \scriptfont1=\seveni  \scriptscriptfont1=\fivei
\textfont2=\tensy \scriptfont2=\sevensy \scriptscriptfont2=\fivesy
\textfont\itfam=\tenit \def\it{\fam\itfam\tenit}
\textfont\bffam=\tenbf \def\bf{\fam\bffam\tenbf} \rm}
%
%
\def\noblackbox{\overfullrule=0pt}
\hyphenation{anom-aly anom-alies coun-ter-term coun-ter-terms}
\def\inv{^{\raise.15ex\hbox{${\scriptscriptstyle -}$}\kern-.05em 1}}
\def\dup{^{\vphantom{1}}}
\def\Dsl{\,\raise.15ex\hbox{/}\mkern-13.5mu D} 
\def\dsl{\raise.15ex\hbox{/}\kern-.57em\partial}
\def\del{\partial}
\def\Psl{\dsl}
\def\tr{{\rm tr}} \def\Tr{{\rm Tr}}
\font\bigit=cmti10 scaled\magstep3
\def\biglie{\hbox{\bigit\$}} 
\def\lspace{}
\def\lbspace{} 
\def\boxeqn#1{\vcenter{\vbox{\hrule\hbox{\vrule\kern3pt\vbox{\kern3pt
        \hbox{${\displaystyle #1}$}\kern3pt}\kern3pt\vrule}\hrule}}}
\def\mbox#1#2{\vcenter{\hrule \hbox{\vrule height#2in
                \kern#1in \vrule} \hrule}}  
\def\tilde{\widetilde} \def\bar{\overline} \def\hat{\widehat}
%
\def\CAG{{\cal A/\cal G}}   
\def\CA{{\cal A}} \def\CC{{\cal C}} \def\CF{{\cal F}} \def\CG{{\cal G}}
\def\CL{{\cal L}} \def\CH{{\cal H}} \def\CI{{\cal I}} \def\CU{{\cal U}}
\def\CB{{\cal B}} \def\CR{{\cal R}} \def\CD{{\cal D}} \def\CT{{\cal T}}
\def\e#1{{\rm e}^{{\textstyle#1}}}
\def\grad#1{\,\nabla\]_{{#1}}\,}
\def\gradgrad#1#2{\,\nabla\]_{{#1}}\nabla\]_{{#2}}\,}
\def\ph{\varphi}
\def\psibar{\overline\psi}
\def\om#1#2{\omega^{#1}{}_{#2}}
\def\vev#1{\langle #1 \rangle}
\def\lform{\hbox{$\sqcup$}\llap{\hbox{$\sqcap$}}}
\def\darr#1{\raise1.5ex\hbox{$\leftrightarrow$}\mkern-16.5mu #1}
\def\lie{\hbox{\it\$}} 
\def\ha{{1\over2}}
\def\half{{\textstyle{1\over2}}} 
\def\roughly#1{\raise.3ex\hbox{$#1$\kern-.75em\lower1ex\hbox{$\sim$}}}
\def\CM{{\cal M}}
\def\CE{{\cal E}}
\def\CI{{\cal I}}
\def\CP{{\cal P}}
\font\fort=cmr10 at 14pt
\font\ninermi=cmti10 at 9pt
\font\ninerm=cmr10 at 9pt
\def\C{{\bf C}}
\def\RR{{\bf R}}
\def\IR{{\bf R}}
%
\def\raca{\titlerm}
\voffset-.4cm
$\,$
\vskip-1.3cm
\rightline{\vbox{\baselineskip12pt\hbox{DFTUZ 91.32}\hbox{FTUAM
91-34}\hbox{(Revised)}}}
\vskip-1.6cm
\Title{}
{\vbox{\centerline{ \raca  Exact
renormalization-group analysis}\vskip2pt\centerline{\raca  of
first order phase transitions{}}\vskip2pt\centerline{\raca
in clock models{}}}}
\centerline{\bf M. Asorey and J.G. Esteve}
\bigskip\baselineskip=12pt plus 2pt minus
1pt\centerline{Departamento de F\'{\i}sica Te\'orica.
 Facultad de Ciencias}
\centerline{Universidad de Zaragoza. 50009 Zaragoza. Spain}\bigskip
\baselineskip=16pt plus 2pt minus 1pt\centerline{\bf J. Salas}
\bigskip\baselineskip=12pt plus 2pt minus
1pt\centerline{Departamento de F\'{\i}sica Te\'orica. C-XI}
\centerline{  Universidad Aut\'onoma de Madrid. Cantoblanco
28049 Madrid. Spain}
\def\rg{renormalization group\ }
\def\rgf{renormalization group flow\ }
\def\rgt{renormalization group transformation\ }
\baselineskip=16pt plus 2pt minus 1pt
\bigskip
\bigskip\medskip

\centerline {\bf Abstract}

We analyze the exact behavior of the  renormalization group
flow in one-dimensional clock-models which undergo first order
phase transitions by the presence of complex interactions.  The
flow, defined by decimation, is shown to be single-valued and
continuous throughout its domain of definition, which contains the
transition points. This fact is  in disagreement with a recently
proposed scenario for first order phase transitions  claiming the
existence of discontinuities of the renormalization group. The
results are in partial agreement with the standard scenario. However
in  the vicinity of some fixed points of the critical surface the
renormalized measure does not correspond to a renormalized
Hamiltonian for some choices of renormalization blocks.  These
pathologies although  similar  to  Griffiths-Pearce pathologies
have  a different physical origin: the complex character of the
interactions. We elucidate the dynamical reason for such a
pathological behavior: entire regions of coupling constants blow up
under the renormalization group transformation.  The flows provide
non-perturbative patterns for  the renormalization group behavior
of electric conductivities in the quantum Hall effect.

\overfullrule=0pt

\hyphenation{systems}

\bigskip\vfill \noindent\baselineskip=12pt plus 2pt minus 1pt
{\bf Keywords}: Renormalization Group, Phase transitions,  Clock
models, Dynamical~systems, \phantom {\bf Keywords:} Quantum Hall
effect. PACS 05.50.+q, 64.60.Fr. 
\Date{ }
\parindent=20pt
\baselineskip=16pt plus 2pt minus 1pt

The behavior of the renormalization group flow around first order
transition points has been a controversial matter for
years. The conventional scenario is based on a {\it smooth}
behavior of the renormalization group flow  and the existence of a
{\it discontinuity fixed point} whose attraction domain contains
the transition surface and has
relevant exponents of the form $y=D$  \ref\rconv{B.
Nienhuis, M. Nauenberg, Phys. Rev. Lett. {\bf 35}, (1975) 477\semi
M. E. Fisher, A.N. Berker, Phys. Rev. {\bf B26}, (1982) 2507}
\ref\rnew {W. Klein, D. J. Wallace, R.K.P. Zia, Phys. Rev. Lett.
{\bf 37}, (1976) 639}, $D$ being the dimensionality of the system.
The singularities associated with first order transitions are
generated by  infinite iterations of renormalization group
transformations in the thermodynamic limit.

However,
recently, some authors claimed to have numerical evidence of the
failure of  the conventional picture \ref\rhas{ H. J. Bl\"ote, R. H.
Swendsen, Phys. Rev. Lett. {\bf 43} (1979) 799\semi C. B. Lang,
Nucl.  Phys.  {\bf B280 }[FS18]
(1987) 255\semi A. Gonz\'alez-Arroyo, M.
Okawa, Y. Shimizu, Phys. Rev. Lett. {\bf 60} (1988) 487}
\ref\rhhas{ A. Hasenfratz, P Hasenfratz,  Nucl.
Phys.  {\bf B295 }[FS21] (1988) 1
\semi K. Decker, A. Hasenfratz, P Hasenfratz, Nucl.
Phys.  {\bf B295 }[FS21] (1988) 21}.  They proposed a new scenario
where
the renormalization group flow is discontinuous at the first order
transition surface. The ambiguity arises because the \rgt associates
to any system on the transition surface as many different
renormalized Gibbs measures as phases coexist at the transition.

More recently, van Enter, Fern\'andez and  Sokal
\ref\sok{ A. C. D. van Enter, R. Fern\'andez and A. D. Sokal, Phys.
Rev. Lett. {\bf 66} (1991) 3253} rigorously proved that the second
picture turns to be false. Their result states that for systems with
bounded
fluctuating variables and absolutely summable real Hamiltonians, the
renormalization group transformation, when properly defined by some
blocking procedure, is single-valued and continuous in a domain
of the space of parameters which includes first order transitions
points.  The only pathology which arises for some   systems (e.g.
Ising model in $D\geq$ 2) is that for some
renormalization group prescriptions (majority rule, decimation,
etc.) the renormalization group transformation is not defined at all
in a neighborhood of  first order transition points, as previously
pointed out by Griffiths-Pearce \ref\rgp{R.B. Griffiths, P.A.
Pearce, Phys. Rev. Lett. {\bf 41} (1978) 917; J. Stat Phys. {\bf 20}
(1979) 449 } and Israel \ref\risr{R. B. Israel, in {\it Random
Fields} Vol. II, eds. J. Fritz, J.L. Lebowitz, D. Sz\'asz, North
Holland, Amsterdam (1981)}. The numerical results of ref.\rhas\
might be understood as artifacts of the truncation of the local
renormalized Hamiltonians used in the Monte Carlo renormalization
group analysis \ref\rgosas {A. Gonz\'alez-Arroyo, J. Salas, Phys.
Lett. {\bf B261} (1991)415 }. The method used in ref. \rhhas\ does
not rely on truncation of local interactions but assumes the
existence of only one relevant perturbation at the fixed point
\ref\rwil{ K. G. Wilson, in {\it Recent Developments in Gauge
Theories}, eds. G.'t et al. Plenum, New York (1980)} and in case of
Ising model there are two: one associated to the temperature
$T=1/\beta$ and another to an external magnetic field $H$
\rnew\rgosas.

In this note we  analyze those issues in some exactly soluble
models recently introduced by two of us \ref\rases{M. Asorey, J.G.
Esteve, J. Stat. Phys. {\bf 65}  (1991) 483}.  The simplicity of
the models allow us to analyze
exact renormalization group flows in finite-dimensional coupling
spaces. Although they exhibit very peculiar properties due
to the presence of complex interactions, we can extract some
rigorous lessons about the behavior of the \rg.  The physical
interest of those models comes from the relevance of their
($\sigma$-model) higher dimensional generalizations in statistical
mechanics  and quantum field theory ($\sigma$-models with similar
complex couplings arise e.g. in the effective description of the
quantum
Hall effect \ref\final{ A. M.
M. Pruisken, Nucl. Phys.  {\bf B 235} (1984) 277\semi
H. Levine, S.
B. Libby,  A. M.
M. Pruisken, Phys. Rev. Lett. {\bf 51} (1983) 1915;
Nucl. Phys.  {\bf B 240} (1984) 30,49,71}). They can also be
considered
as one-dimensional versions of Chiral Potts models which have been a
recent focus of interest \ref\chiralpotts1{
H. Au-Yang, B. M. McCoy, J. H. H. Perk, S. Tang and M. L. Young,
{ Phys. Lett.} {\bf A 123} (1987) 219}.

The complex character of the interactions of  the models makes
possible the existence of first order phase transitions though
they have only short range interactions (in fact nearest-neighbor),
which is impossible for one-dimensional systems with real
Hamiltonians (see ref. \ref\G{J. O. Georgii, {\it
Gibbs Measures and Phase Transitions}, de Gruyter, Berlin-New York
(1988)} for a recent review). Another unusual property of the
models is that the correlation length at the (first-order)
transition surface is infinite which implies  a very peculiar
behavior of finite size effects in the vicinity of the  critical
points \ref\rnew{M. Asorey, J.G. Esteve, J. Salas, DFTUZ 93.3}.

 Our results for the three and four-state clock models are
in agreement with the conventional  scenario for the renormalization
group  near first order critical points. The \rgt is single-valued
and continuous throughout its domain of definition which contains
 transition points, and has discontinuity fixed points with the
peculiarity  that the correlation length at those  fixed points is
infinite ($\xi=\infty$).

However, there are some pathologies  of the systems at the points
of the
coupling space where the \rgt is not defined. They arise because of
the appearance of  renormalized measures without renormalized
Hamiltonians and  by no means  are related to the
ambiguities of the discontinuous scenario.
In these sense they are analogous to
Griffiths-Pearce singularities. In the present case there
is also  a dynamical reason behind such  a pathological behavior:
the blow up of the corresponding renormalization group trajectories
at infinite under a finite number of (fractional) iterations. The
renormalization group flow generates a (local) one-parameter group
of local transformations \ref\kob {S. Kobayashi, K. Nomizu, {\it
Foundations of Differential Geometry}, vol. 1, Interscience, New
York (1963)} but there is no global renormalization
 group. If we exclude the pathological region of
parameters in the vicinity of first order transition points the
renormalization group is globally defined. This standard dynamical
behavior is the reason for such a pathological behavior.
However, the physical origin of these pathologies is  different from
that of Griffiths-Pearce singularities. In this case is the existence
of complex interactions interactions which makes possible the
appearance of  renormalized measures which vanish for a
very large sets of configurations and cannot be described in
terms of  renormalized Hamiltonians.
In the present models the pathological values of the couplings are
non-generic and  non universal. They depend on the renormalization
group prescription and  might be thought as   artifacts of the choice
of the  prescription. However, we find pathologies for any
prescription based on decimation.

 The renormalization of systems with complex topological
$\theta$-terms ($\sigma$-models, gauge theories, etc) has been always
problematic from the numerical point of view, and semiclassical
considerations based in instanton calculus cannot be taken for grant.
The exact results obtained in our models shed some light in the
possible
(pathological) behavior of the renormalization group for those
systems
and provide a non-perturbative pattern which seems to pick up the
essential behavior of $\theta$-terms under renormalization.

We consider the $q$-state clock of a classical spin variable $\vec
s_n$  fluctuating  among the  $q$-roots of unity,
\eqn\uno{\vec s_n=(\cos {  2  \pi p_n \over q},
\sin {2 \pi p_n \over
q}) \qquad p_n=0,\dots,q-1}
in one-dimensional space with
interacting Hamiltonian \rases\
\eqn\euno{\eqalign{\beta{{\cal{H}}}&=-\sum_{n=1}^N \left( J \vec s_n
\cdot\vec s_{n+1} - i\vec{\varepsilon}\cdot (\vec s_n\times \vec
s_{n+1})\right)\cr
&=-\sum_{n=1}^N \left(
J\cos{  2  \pi  \over q}(p_n-p_{n+1})+
i\varepsilon\sin{  2  \pi  \over q}(p_n-p_{n+1})\right),\cr}}
where $\beta$ denotes as usual the inverse of the temperature.
The  eigenvalues of the transfer matrix $T$
\eqn\edos{\lambda_k=\sum_{n=0}^{q-1}{\rm exp }\ \{{ J\cos(2\pi
n/q)+i \varepsilon \sin(2\pi n/q)-i k (2\pi n/q)}\}}
cross each other an infinity number of times. The crossings of
leading levels correspond to first order transition points with
infinite correlation length \rases. In the case of 3-state clock
model the  three eigenvalues of the transfer matrix
\eqn\etres{\lambda_0={\rm e}^{J}+2{\rm
e}^{-{J\over 2}}\cos (\theta),\quad\lambda_1={\rm e}^{J}-2{\rm
e}^{-{J\over 2}}\cos (\theta+\pi/3),\quad
\lambda_2={\rm e}^{ J}-2{\rm
e}^{-{J\over 2}}\cos(\theta-\pi/3)}
have leading level crossings at the transition temperatures
$\theta=(2m+1){\pi/ 3}$,
where  $\theta =\varepsilon\sqrt{3}/2$.
Single alternate decimation induces the following renormalization
group transformation
\eqn\ecinco{J'={2\over 3} \log { {\rm e}^{2J}+
2{\rm e}^{-
J}\over ({\rm e}^{-2J}+ 4{\rm e}^{
J}+4{\rm e}^{-J\over 2}\cos{3 \theta})^{\ha}},\quad\theta'=\arctan
{2{\rm e}^{J\over 2}\sin{\theta}-{\rm
e}^{-J}\sin{2\theta}\over 2{\rm e}^{J\over 2}\cos{\theta}+{\rm
e}^{-J}\cos{2\theta}}.}
There is an additional free energy renormalization which is not
relevant for the analysis of the critical behavior of correlation
functions. The transformation \ecinco\ has to be understood as the
projection of the renormalization group transformation on the plane
of coupling constants $(J,\theta)$.

The flow is univocally  defined everywhere and exhibits a very
interesting dynamical behavior. There are not traces of the
ambiguities associated to discontinuous scenarios. The critical
surfaces $\theta={(2m+1)\pi/3}$ are invariant under renormalization
group transformations as well as the subcritical ones $\theta= 2
{m\pi/3}$. There are three kinds of fixed points (see Fig. 1)
$${\vcenter{\vbox{\hrule\hbox
{\vrule\kern3pt\vbox{\kern3pt
\halign
{#\hfil&\qquad\hfil #\hfil &\qquad #\hfil\cr
$F^n_\infty \qquad
J=0,\theta=2n\pi/3$&{\rm (stable)}\
&{\rm infinite\  temperature}\cr
$F^\theta_0 \qquad\, J=\infty$ &{\rm (unstable)}
&{\rm zero\  temperature}\cr
$F_c^n \qquad J={2\over
3}\log 2,{ \theta=(2n+1)\pi/3}$ &{\rm (saddle\
point)} &{\rm transition\
temperature}\cr}\kern3pt}\kern3pt\vrule}\hrule}}}
$$
The third class of fixed points $F_c^n$ are sitting on the transition
curves and attract all transition points. There is one relevant
direction at those points which is tangent to the renormalized
trajectory
\eqn\esiete{\theta= \arccos( {{\rm e}^{3J/2}\over 2}) +
(2m+1){\pi\over
3}}
and flows towards the infinite temperature fixed points
$F^n_\infty$.
The correlation length vanishes at the infinite temperature
fixed point $F^n_\infty $ and becomes infinite at the  points
$F^n_c $ and $F^\theta_0 $ of the critical surface.

The only second order critical points where the continuum limit can
be obtained by scaling are the zero temperature fixed points
$F^\theta_0$.
The existence of a line of fixed points leads to the
existence of different quantum mechanical systems \ref\rrotor{M.
Asorey, J.G. Esteve, A. F. Pacheco, Phys. Rev {\bf 27} (1983)
1852}.

Another interesting aspect of the renormalization group
are the critical exponents. The linearized renormalization group
tranformations around the different fixed points are
\eqn\eocho{L_{F^n_\infty}=\pmatrix{0
&0\cr 0 &0\cr}\quad L_{F^n_c}=\pmatrix{0 &0\cr 0
&2\cr}\quad L_{F^\theta_0}=\pmatrix{1 &0\cr 0 &1\cr}.}
Therefore, the infinite temperature fixed points ${F^n_\infty}$ are
stable and have two irrelevant  directions
($\lambda_i=2^{y_i}=0, i=1,2$).

The second order transition points at zero temperature have two
marginal perturbations ($y=0$), one along the line of fixed points
$J=0$ and another tangent to a
renormalized trajectory flowing away towards the other types of
fixed points.

The fixed points  on the first order transition surface ${F^n_c}$
are saddle points with one relevant direction
($\lambda_1=2^{y_1}=2$)
flowing along the renormalized trajectory \esiete\ towards the
the infinite temperature fixed points ${F^n_\infty}$ and
${F^{n+1}_\infty}$ and one irrelevant direction
($\lambda_2=2^{y_2}=0$) along the critical line $\theta
=(2n+1)\pi/3$.
The existence of one relevant perturbation with
$y=1$ agrees with
the picture advocated by the standard scenario for
discontinuity fixed points of first order transitions except for
the fact that the correlation length is infinite. In general, the
number of relevant perturbations with critical dimension equal to
the space dimension $D=1$ has be equal to the number of phases minus
one.
In this case
it implies that the critical exponent $\nu =1$.

The only amazing behavior of the \rgf arises for
\eqn\enueve{J< J_0(\theta)={2\over 3} \log 2\cos
(\theta-(2m+1){\pi\over 3}).}
where some eigenvalues of the transfer matrix $T$ become negative.
One single iteration of the \rgt maps any point  of this region
into a point of the    region $J > J_0(\theta)$ where the transfer
matrix
is positive ($T>0$),
and the \rgt is singular at the transition points $J=- {2\over 3}
\log 2, \theta={(2 n+1)\pi/ 3}$ which  are mapped by \ecinco\ into
$J=\infty , \theta={(2 n+1)\pi/ 3}$.
Since
our decimation procedure transforms the transfer
matrix $T$ into $T^2$, which  is always  positive the corresponding
\rgt  maps the points of the region $J<J_0(\theta)$ into those of the
region
$J>J_0(\theta)$. The renormalized trajectory \esiete\ is the
borderline between those domains.

The singularity at the points $J=\infty , \theta={(2 n+1)\pi/ 3}$ is
due to the diagonal character of  $T^2$. The  vanishing of
non-diagonal entries
implies that the renormalized measure is not Gibbsian and does not
correspond to any regular renormalized Hamiltonian.

In order
to analyze  a more natural flow for Hamiltonians with $T\not\geq 0$
we
define a \rgt by double decimation (see Fig. 2), which leaves
invariant the domain where $T$ is not non-positive. The
corresponding \rgt
\eqn\eonce{\eqalign{&J''={2\over 3}\log{{\rm e}^{3J}+6+2 {\rm
e}^{-3J/2} \cos 3\theta\over 3(4 \cosh^2 3J/2 + 1+4\cosh 3J/2\cos
3\theta)^\ha}\cr  &\theta''= \arctan{2\cosh 3J/2\sin \theta-\sin
2\theta\over 2\cosh 3J/2\cos \theta +\cos 2\theta}\cr}}
exhibits a similar global ({\it
universal}) behavior for the $T>0$ region $J>
J_0(\theta) $ (see Fig. 1), and it is
not singular at $J=- {2\over 3} \log 2, \theta={(2 n+1)\pi/ 3}$. In
fact, these points become
fixed (unstable) points. However, the transformation \eonce\
is again singular at the points satisfying that
\eqn\enueva{{\rm e}^{3J}+6+2 {\rm
e}^{-3J/2} \cos 3\theta=0,}
which are mapped into the infinite line $J=-\infty$.

The analysis of the \rgf  for ${\rm e}^{3J}+6+2 {\rm
e}^{-3J/2} \cos 3\theta<0$ requires an analytic extension of the
model to the  complex plane of the coupling constant $J$
\ref\aefs{M. Asorey, J.G. Esteve, R. Fernandez, J. Salas, Nucl.
Phys. {\bf B} (in press)}. The only really pathological points are
those of the curve \enueva\ which can be characterized by the fact
that  the { diagonal} elements of $T^3$  vanish.  In  such a case the
renormalized measure  vanishes for a large set of clock variables
configurations and cannot be described in terms of Gibbsian positive
weights
associated to a renormalized Hamiltonian.

Therefore, the
pathological behavior of systems without non-negative transfer matrix
can be improved but not  completely cured by this
change of  the blocking procedure. However, we remark that the set
of pathological Hamiltonians is non-generic in the space of
couplings constants: it has codimension one or two depending on the
blocking procedure.

 The universal behavior for
$J> J_0(\theta)$ also
holds for the continuous flow associated to the transformation
$T\rightarrow T^t$ which is  only valid for $T>0$. The integration of
the corresponding differential equations
\eqn\edoce{\eqalign{&\dot {J}= {2\over 3}{\rm e}^{-J} \lambda_0
\log \lambda_0+ {2\over 3}(1+{2{\rm e}^{-3J/2}\over \cos \theta})
{\cot\theta} \dot {\theta}- {{\rm e}^{J/2}+2{\rm e}^{-J}\cos
\theta \over 3\sqrt {3} \sin \theta} (\lambda _1\log
\lambda_1-\lambda _2\log \lambda_2)\cr &{\dot \theta} =
{\sqrt{3}\over 18 } {\rm e}^{J/2}(3 \cos \theta (\lambda_1\log
\lambda_1- \lambda_2\log \lambda_2)- \sqrt{3} \sin \theta
(2\lambda_0\log \lambda_0 - \lambda_1\log
\lambda_1-\lambda_2\log \lambda_2))\cr}}
yields a continuous

flow with the same global characteristics  and the same picture for
first order transition points (see Fig. 1).

The behavior of the renormalization group is essentially changed by
the presence of a external real magnetic  field $H$. When $H\neq 0$
there are not first order transition points  and Lee-Yang theorem
holds in this case. The renormalization group can be exactly
analyzed  by introducing a new coupling which breaks the internal
rotation symmetry of the clock model. The only first order critical
points $F_c$ belong to the surface $H=0$ and a new relevant direction
appears at $F^n_c$ along the renormalized trajectory associated to
the external magnetic perturbation.

In the case of the 4-state model whose critical behavior was
analyzed in \rases, the \rgt induced by decimation is not well
defined unless we introduce a new interaction term of the form
$J_1 \cos {\pi}(p_n-p_{n+1})$ in the
Hamiltonian.
Therefore we need to consider a new family of models with
Hamiltonians
\eqn\etrece
{\eqalign{\beta{{\cal{H'}}}&= -\sum_{n=1}^N
\left(
J\cos{   \pi  \over 2}(p_n-p_{n+1})+
J_1 \cos {\pi
}(p_n-p_{n+1})+
i\varepsilon\sin{   \pi  \over 2}(p_n-p_{n+1})\right).\cr}}
The  eigenvalues of the transfer matrix are given by
\eqn\ecatorce{\eqalign{&\lambda_0=2{\rm e}^{J_1}\cosh J +
2{\rm e}^{-J_1}\cos \varepsilon\qquad\lambda_1=2{\rm e}^{J_1}
\sinh J +
2{\rm e}^{-J_1}\sin \varepsilon\cr
&\lambda_2=2{\rm e}^{J_1}\cosh J -
2{\rm e}^{-J_1}\cos \varepsilon\qquad\lambda_3=2{\rm e}^{J_1}\sinh J
-2{\rm e}^{-J_1}\sin \varepsilon \cr}}
and crossings of leading eigenvalues correspond again to    first
order transition points with infinite correlation length.

An alternate single decimation defines the following \rgt
\eqn\equince{\eqalign{J'= &{1\over 2} \log { 1+{\rm e}^{4J_1}\cosh
{2J}\over {\rm e}^{4J_1}+ \cos{2 \varepsilon}}\cr
J_1'=&{1\over 4}\log {(1+{\rm e}^{-4J_1}\cos {2\varepsilon})(1
+{\rm e}^{4J_1}\cosh {2J})\over
2\cosh {2J}+2\cos{2\varepsilon}}\cr
\varepsilon'=&{1\over 2}\arccos {1+\cosh {2J}\cos{2\varepsilon}\over
\cosh {2J}+\cos{2\varepsilon}}.\cr
}}
In this case a new problem arises. The transformation \equince\ is
not defined for couplings of the domain
\eqn\edieciseis{B_2=\{(J,J_1,\varepsilon);
{\rm e}^{4J_1}<-\cos 2\varepsilon ,\
\varepsilon>{\pi\over 4}\}, \qquad \qquad \qquad {\rm (Black\,\,
Hole)}}
in terms of  renormalized hamiltonians  of the type \etrece\ with
real couplings $J,J_1,\epsilon$. However, there exist an analytic
extension of ${\cal{H'}}$ to the complex planes of $J$ and $J_1$
that preserves the hermitian character of the transfer matrix; and
the renormalized hamiltonians  of
the points of $B_2$ belongs to such a class of hamiltonians \aefs.
Thus, the only real pathology arises for couplings in the border of
the black hole
$$ \partial B_2=\{(J,J_1,\varepsilon);
{\rm e}^{4J_1}=-\cos 2\varepsilon ,\
\varepsilon>{\pi\over 4}\}.$$
Those pathological points are non-generic and  vary with the
definition of the renormalizacion group transformation.  A common
characteristic of those points is that they yield  renormalized
measures which vanish for a large
set of clock variables configurations and cannot be described in
terms of a Gibbs measure associated to a renormalized Hamiltonian.

Apart from such a
pathological zone the \rg is well defined everywhere and none
ambiguity arises at first order transition surfaces. Because of the
periodicity in $\varepsilon$ it suffices to consider the region
$0\leq\varepsilon\leq{\pi\over 2}$. The critical surfaces in that
region are
\eqna\ediecisiete
$$\eqalignno{&{\rm e}^{-J+2 J_1}=-\sqrt{2}
\cos(\varepsilon+{\pi\over 4})\qquad\qquad {\pi\over 4
}<\varepsilon<{\pi\over
2 }\ &\ediecisiete {a}\cr
&J< 2J_1,\quad
\varepsilon={\pi\over 2} & \ediecisiete {b} }$$
and the \rg  fixed points are listed below (see also Fig. 3)
$${\vcenter{\vbox{\hrule\hbox
{\vrule\kern3pt\vbox{\kern3pt
\halign
{#\hfil&\quad#\hfil&\quad\hfil #\hfil &\qquad #\hfil\cr
$F_\infty
$&$J_1=J=\varepsilon=0$
&{\rm (stable)}\
&{\rm infinite\  temperature}\cr
$F_0 $  &$J=J_1=\infty$ &{\rm (unstable)}
&{\rm zero\  temperature}\cr
$F^\varepsilon_0 $
&$J=\infty,J_1={1\over
4}\log\cos 2\varepsilon,0<\varepsilon<{\pi/4}$ &{\rm (saddle\
point)} &{\rm zero\
temperature}\cr
$F_c $ &$ J={1\over
2}\log 3,J_1={1\over
4}\log 3, { \varepsilon=\pi/2}$ &{\rm (saddle\
point)} &{\rm transition\
temperature}\cr
$F^\ast_c $ &$ J=\infty ,J_1=-\infty,
{ \varepsilon=\pi/4}
$ &{\rm (saddle\
point)} &{\rm zero\
temperature}\cr
$F^\dagger_c $&$ J=0 ,J_1=\infty,
{ \varepsilon=\pi/2}
$ &{\rm (saddle\
point)} &{\rm zero\
temperature}
\cr
$F^\dagger $&$ J=0 ,J_1=\infty, {
\varepsilon =0}
$ &{\rm (saddle\
point)} &{\rm zero\
temperature}\cr
}\kern3pt}\kern3pt\vrule}\hrule}}}
$$

Once again  positivity provides a useful
information about the renormalization group flow.
 The domain where   the transfer matrix is positive ($T>0$) is
located below the surface
\eqn\ediecinueve{J_0 (J_1,\varepsilon) =\max\{ {\rm arcosh }\ ({\rm
e}^{-2J_1}\cos \varepsilon), {\rm arcsinh }\ ({\rm e}^{-2J_1}\sin
\varepsilon)\} .}
The points above such a surface $J<J_0  (J_1,\varepsilon)$ where
$T$ is not non-positive are mapped by one single iteration of the
\rgt into points below the surface, $J>J_0  (J_1,\varepsilon)$,
 by the same reason that in the case of the
3-state clock model. A double iteration would cure this anomaly but
then the
 domain of definition of the \rgt in terms of Hamiltonians of
the type \etrece\ with
real couplings $J,J_1,\epsilon$ is smaller than that of  a single
alternate  decimation \equince.
 The \rgf
 for the region where $T$ is not non-positive  $J< J_0 $ will be
discussed in \aefs. Let us concentrate here on the
 region with positive transfer matrix $J> J_0 $.
On the first order transition surface the only fixed points are
$F_c$,  $F^\ast_c$ and $F^\dagger_c$. Although the attraction
domain of  all those points  contains transition points, only $F_c$
has one eigenvalue $2\,\, (y=1)$ of the linearized \rgt  as required
in
the
standard Nienhuis-Nauenberg picture
 for discontinuity fixed points. In our case it implies that the
 critical exponent $\nu =1$,

 The renormalization group   is well defined
on the domain of the transition surface with $J< J_0 $.  Transition
points with $J<2J_1$
 in the plane $\varepsilon=\pi/2$     are attracted by
$F^\dagger_c$,  transition points with  $J=2J_1$ move towards the
fixed point $F^\ast_c$ along the renormalized trajectory defined by
the edge \ $J=2J_1, \varepsilon=\pi/2$ of the transition surfaces
\ediecisiete {a}\
and \ediecisiete {b}, and the
remainder   points   of the transition surface  with  $J>2J_1$ flow
towards the fixed point $F^\ast_c$.  A similar behavior is observed
for
points below the
transition surface with  $J>2J_1$, which reach  the black hole domain
$B_2$ after a finite number of iterations.  This is one of the
reasons why the
 (complex) region  $B_{2n}$  (black hole) for $T^{2n}$ grows with
$n$. The new pathological points for $T^{2n}$ are points that
already  reached the region $B_2$ in less than n iterations of $T^2$.

On the other hand, the
continuous flow defined for  systems with  $T>0$ by the
equations
\eqn\eveinte{\eqalign{&\dot {J}=
 {1\over
4}(\lambda_1\log
\lambda_1 + \lambda_3\log \lambda_3){\rm e}^{J_1}\cosh J  -
{1\over
4}(\lambda_0\log \lambda_0 + \lambda_2\log \lambda_2){\rm
e}^{J_1}\sinh
J    \cr
&\dot {J}_1=
 { (\lambda_0\log
\lambda_0 + \lambda_2\log \lambda_2)\over {8{\rm e}^{J_1}
  }  \cosh J}- {(\lambda_0\log \lambda_0 - \lambda_2\log
\lambda_2)\over{8{\rm e}^{-J_1}   } \cos \varepsilon }-
{\dot \varepsilon}{\tan
\varepsilon\over 2}-\dot {J}{\tanh J\over 2}\cr
 &{\dot \varepsilon} = {{\rm e}^{J_1}\over 4 } (
\cos \varepsilon (\lambda_1\log \lambda_1- \lambda_3\log
\lambda_3)-
\sin \varepsilon  (\lambda_0\log \lambda_0 - \lambda_2\log
\lambda_2))\cr}}
is defined everywhere below the surface \ediecinueve. In
particular, it is well defined in the pathological points of
$\partial B_2$.   The only pathological property of the points
of the domain $B_2$ is that they have been swallowed up by  the line
$J=\infty, \ J_1=-\infty , \varepsilon \in (\pi/4,\pi/2]$
in a finite {\it time} $t<2$. This
feature
explains why there
is a restriction on the  domain of definition of $T^2$.
In fact,
all the points below the transition surface will reach such a line
in a finite time, which means that any given point of such a domain
does not belong to the domain of $T^n$ for $n$ large enough.
The points at the transition surface \ediecisiete{a}\ with $J>2
J_1,\  J>J_0$ are also attracted by the  fixed
point  $F^\ast_c$ but in  an infinite time (The
intersection curve of the critical surface and $J=J_0$
 is the renormalized trajectory joining
$F^\ast_c$ and  $F_c$ (see Fig. 3)). Therefore, the
renormalization group transformation is well defined in such points
for any kind of decimation blocking.

 The pathologies observed so far in the  definition of
the \rgt can   be then understood as a pure consequence of the
dynamical behavior of a \rgf with a very strong attractive domain.
For any continuous flow with such a strong domain attractor the
corresponding ({\it local}) one-parameter group of local
transformations cannot be implemented to a global group of
transformations \kob. In the present case the transformation $T^t$
cannot be globally defined for any value of $t\neq 0$. However, if we
exclude the region below the transition surface \ediecisiete{a}\ the
\rgf defines a global one-parameter group of transformations for any
value of $t\in \IR$.

In summary, besides the above dynamical system explanation,
we have shown three sources of pathological behavior of the
renormalization group, all of them motivated by the complex nature of
the
interacting terms of the Hamiltonian:

i) The presence of negative eigenvalues in the transfer matrix yields
an odd
behaviour under decimation renormalization group transformations
which
generates pathologies. In particular, a continuum renormalization
group cannot be implemented in that case.

ii) In the 4-states model there are some values of the parameters
(region
 $B_2$) where the
transfer matrix is positive but the renormalization group is not
defined.
This can happen
because the correspondence between eigenvalues of $T^2$ and the
parameters
of the renormalized Hamiltonian is not analytic.

iii) It is
possible to extend the domain of definition of the \rgt by analytic
continuation in the space of coupling parameters \aefs.
In this way the set of pathological Hamiltonians can be reduced to
 non-generic hypersurfaces with non-trivial codimension in the space
of coupling parameters (e.g. the boundary of $B_2$).

In any case,  although the Hamiltonians of these models are complex
the
 Boltzmann
weights can be paired to give rise to real contributions. The main
difference
 with standard  systems with real couplings is that some of the
weights might
vanish. However, the
configurations with vanishing contributions are non-generic and
should not be confused with the pathological renormalized measures,
where the configurations
with vanishing contributions become generic (e.g. the transfer matrix
has null
all non-diagonal entries).

 One might conjecture that the pathological behavior of \rgt
 observed in higher dimensional systems is
generated by similar
dynamical properties. In fact, the same behavior occurs in the \rgf
of the $q$-state clock models with Hamiltonians
\eqn\efinal{\beta {{\cal{H'}}}=-\sum_{n=1,r=1}^{N,[{q/ 2}]}
\left( J_r \cos{  2 r \pi  \over q}(p_n-p_{n+1})+
i\varepsilon_r\sin{  2 r \pi  \over q}(p_n-p_{n+1})\right)}
which are the natural generalization of the systems \euno.

 Finally,  we remark that the \rgf of the $q=3$ model (Fig. 1) is
similar to the expected flow of  dissipative
$\sigma_{xx}$ and Hall $\sigma_{xy}$ conductivities in the quantum
Hall effect when described in terms of a $\sigma$-model \ref\Hall{D.
E.
Khmel'nitskii, Sov.
Phys. JEPT Lett. {\bf 38} (1983) 552\semi V. K. Knizhnik,
A. Yu. Morozov, Sov.
Phys. JEPT Lett. {\bf 39} (1984) 240 \semi H. Levine, S.
B. Libby, Phys. Lett. {\bf B 150} (1985) 182 \semi A. M.
M. Pruisken, Phys. Rev. {\bf B 15} (1985) 2636 }.
Although the clock model  can be considered as a one-dimensional
discretization of such a $\sigma$-model, it was not expected to
 belong to the
same renormalization class and provide an exact non-perturbative
pattern for  the renormalization group behavior of those
conductivities (We thank  Alexei Morozov for pointing out such a
connection).

The most remarkable property of this pattern is that in the
thermondynamic limit the dissipative conductivity would vanish and
the
Hall conductivity would be quantized as it is observed in the quantum
Hall effect (see fig.1 of the first and third papers of Ref. \Hall).
 These facts raise the conjecture that a solution of the
strong CP-problem of QCD might be
found in a similar way in terms of a renormalization flow of the
$\theta$
parameter towards quantized  (unobservable) values $\theta = 2\pi n$.
In the continuum limit all values of $\theta$ are consistent but in a
given (perhaps fundamental) discretisation the topological
CP-breaking
couplings flow in the infrared to the quantized CP-preserving values
$2\pi n$.

        \bigbreak\bigskip\bigskip\centerline{{\bf
Acknowledgements}}\nobreak
 We thank Roberto Fern\' andez, Antonio  Gonz\' alez-Arroyo and
Alexei Morozov for illuminating discussions.
 We also thank  Prof. Michael E.  Fisher
 for correspondence. J. S. thanks  the members of Departamento de
F\'{\i}sica Te\' orica of  Universidad de Zaragoza  for their
hospitality during the completion of this work. We also acknowledge
CICYT for partial finantial support (grants AEN90-0029,
PB90-0916).

 \listrefs

\figures
\fig {1.} {Renormalization group flow for the 3-state model. The
renormalization group transformation is universal for $J>
J_0 (J_1,\varepsilon) $ ($T>0$). The renormalization group
trajectories are integral curves of  the continuous flow defined by
equation  \edoce. Besides the zero temperature fixed points
$F^\theta_0$, which are not displayed in the picture, there are
two types of fixed points:  infinite temperature fixed points
$F_\infty^n\, (J=0, \theta=2n\pi/3)$,   and  discontinuity fixed
points
$F_c^n\, (J={2\over 3} \log 2,\, \theta=(2n+1)\pi/3)$. The later
behave as  attractors for the  transition points on the lines
$\theta=
(2n+1) \pi/3$. In the  domain $J<
J_0 (J_1,\varepsilon)
$ where $T<0$,the
renormalization group is non-universal. In fact, it becomes singular
at the points $F_n$ with $ J=-2/3\ \log 2,\, \theta=(2n+1)\pi/3$ and
the curve ${\rm e}^{3J}+6+2 {\rm e}^{-3J/2} \cos 3\theta=0$ (dashed
curve) for  single  and double decimations, respectively.}
\fig {2.}
Single and double decimations.
\fig {3.}
{Renormalization group flow for 4-state model.
  Shadowed surfaces correspond to transition points satisfying the
 positivity condition $T>0$. For simplicity, we only show the
spectrum of fixed points and renormalized trajectories of the
renormalization group transformation. The boundary of the domain
$B_2$, $\partial B_2=\{(J,J_1,\varepsilon),
{\rm e}^{4J_1}=-\cos 2\varepsilon ,\
\varepsilon>{\pi\over 4}\}$
where the \rgt is not defined can be seen on the  left back
corner of the $(\epsilon,e^{-4J_1},0)$ plane.}

\end